# Local Interstellar Hydrogen's Disappearance at 1AU: Four Years of IBEX in the rising solar cycle.

Lukas Saul[1], Maciej Bzowski[3], Stephen Fuselier[4], Marzena Kubiak[3], Dave McComas[4,5], Eberhard Möbius[2,6], Justina Sokół[3], Diego Rodriguez[1], Juergen Scheer[1], Peter Wurz[1]

(1) University of Bern, Bern, Switzerland
(2) University of New Hampshire, Durham NH, USA
(3) Space Research Centre PAS, Warsaw, Poland
(4) Southwest Research Institute, San Antonio TX, USA
(5) University of Texas at San Antonio, San Antonio TX, USA
(6) Los Alamos National Laboratory, Los Alamos, NM, USA

**Abstract.** NASA's Interstellar Boundary Explorer mission has recently opened a new window on the interstellar medium by imaging neutral atoms. One "bright" feature in the sky is the interstellar wind flowing into the solar system. Composed of remnants of stellar explosions as well as primordial gas and plasma, the interstellar medium is by no means uniform. The interaction of the local interstellar medium with the solar wind shapes our heliospheric environment with hydrogen being the dominant component of the very local interstellar medium. In this paper we report on direct sampling of the neutral hydrogen of the local interstellar medium over four years of IBEX observations. The hydrogen wind observed at 1 AU has decreased and nearly disappeared as the solar activity has increased over the last four years; the signal at 1 AU has dropped off in 2012 by a factor of ~8 to near background levels. The longitudinal offset has also increased with time presumably due to greater radiation pressure deflecting the interstellar wind. We present longitudinal and latitudinal arrival direction measurements of the bulk flow as measured over four years beginning at near solar minimum conditions. The H distribution we observe at 1AU is expected to be different from that outside the heliopause due to ionization, photon pressure, gravity, and filtration by interactions with heliospheric plasma populations. These observations provide an important benchmark for modeling of the global heliospheric interaction. Based on these observations we suggest a further course of scientific action to observe neutral Hydrogen over a full solar cycle with IBEX.

## 1. IBEX-Lo

NASA's Interstellar Boundary Explorer is a small explorer class mission (SMEX) consisting of an Earth orbiting satellite launched in the fall of 2008. The IBEX spacecraft is on a highly elliptic orbit and has two energetic neutral atom sensors covering overlapping energy ranges [McComas et al., 2009]. The interstellar neutral wind is in the energy range of the IBEX-Lo sensor [Fuselier et al., 2009; Möbius et al., 2009]. To observe and characterize neutral atoms in this energy range a surface conversion system is necessary because the atoms are not energetic enough to pass through the thin foils used in most time-of-flight (TOF) mass spectrometers [Wurz, 2000]. Surfaces suitable for such neutral to negative ion conversion were tested at the University of Bern in Switzerland and are described by Scheer et al. [2006] and Wurz et al., [2006]. The surface chosen for the IBEX-Lo instrument was chemical vapor deposited (CVD) diamond-like carbon. This material is pure carbon, however at the surface where there are open chemical bonds it is hydrogen terminated, and so energetic neutral atoms incident on the surface produce a signal of sputtered hydrogen ions. While this is an important feature for the detection of noble gases such as helium, it presents a background for our hydrogen measurement. The backgrounds present in the IBEX-Lo measurements from various sources are described by Wurz et al. [2009].

IBEX-Lo is designed to maximize the observed signal of ENAs by using a large geometric factor, which enables mass-, energy-, and angle-resolved measurements and the ability to measure very small fluxes. The IBEX-Lo sensor incorporates a number of innovations that optimize the signal to noise ratio and reduce the background count rate. Most importantly, IBEX-Lo, like IBEX-Hi [Funsten et al., 2009], is based on a "Bundt Pan" type electrostatic analyzer, which focuses a very large circular aperture onto a much smaller, central mass spectrometer, providing a largely enhanced signal to noise [Fuselier et al., 2009; McComas et al., 2009]. In addition the mass spectrometer in IBEX-Lo is a TOF system that uses a triple coincidence detection system, which measures the time of flight of an atom in the mass spectrometer three ways to verify a valid detection and eliminate background. For the purposes of this paper we only use fully qualified triple coincidence counts, i.e. counts in which all three TOF measurements agree for the mass of the particle, to reduce background rates as described by Wurz et al. (2009).

To determine the flux of interstellar hydrogen from the local interstellar medium (ISM) entering the IBEX-Lo sensor and converted to negative ions on the conversion surface, it is necessary to separate the background hydrogen due to surface sputtering by ISM helium. Two approaches to statistically separating the converted hydrogen component from the sputtered component are outlined in Saul et al. [2012]. For this work we use the energy analysis method to separate the converted interstellar hydrogen component. This method relies on the difference of the energy spectrum for hydrogen produced by sputtering from that of converted Hydrogen [Möbius et al., 2012]. The raw energy spectra taken during the interstellar Helium peak as well as the interstellar Hydrogen peak are shown in Figure 1.

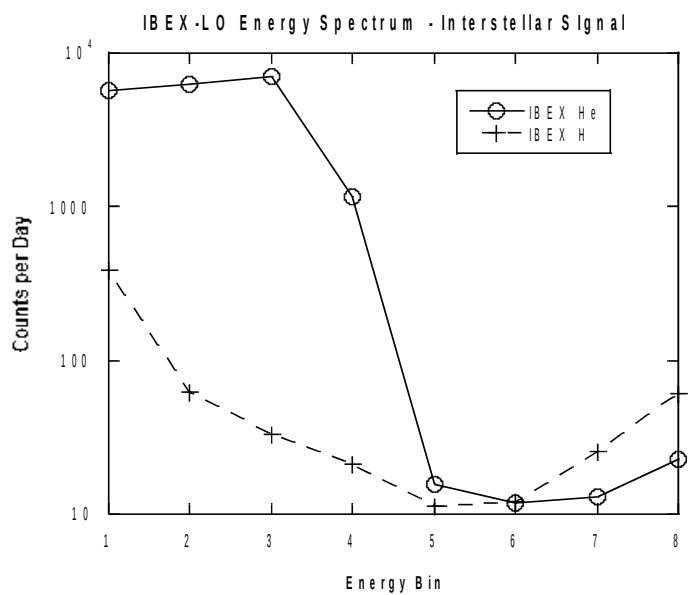

Figure 1. A comparison of IBEX-Lo sensor Hydrogen, during one orbit produced by interstellar helium wind sputtering (orbit 18) and in the other mostly by converted interstellar hydrogen wind (orbit 23). Shown are raw triple coincidence $H^0$ events, which come from the 1/10 of spin phase range (36 degrees) bracketing the maximum interstellar flow.

The energy spectrum of the sputtered component taken during the peak of interstellar He wind observation matches very well to the calibration measurements taken at the University of Bern using a low energy neutral Helium beam to simulate the incoming interstellar matter. In contrast to the He wind observation we can see that the H interstellar signal is peaked at the lowest energy bin, with some flux in the second bin.. The increase in flux in bins 7 and 8 is due to omnidirectional background of high energy ions which increases during the times when the spacecraft is in the Earth's magnetotail and is unrelated to the interstellar flow.

We found that the results using the TOF spectrum to separate the species are generally the same as for using this energy method [Saul et al., 2012]. However, we caution that improvements need be made in this methodology in order to carry out detailed analysis of the low-energy wings of the hydrogen distribution. In particular, our method for separation relies on certain assumptions about the incoming energy of the interstellar helium and hydrogen. In fact, there are unique distributions of energies for each species. A more complete analysis will take this into account. For this paper we use a simple energy analysis as outlined in our previous paper; this method is adequate for determining bulk parameters.

## 2. Local Interstellar Medium

By observing absorption lines in the light of nearby stars one can determine many line-of-sight integrated properties of the surrounding interstellar medium [Frisch et al., 2002; Redfield & Linsky 2008, Frisch et al., 2011]. Some indirect local measurements of this material have also been made. Solar radiation scatters from the neutral component of the interstellar gas, which has been observed for hydrogen [Lallement et al., 2004] and helium [Vallerga et al., 2004; Chassefiere et al., 1986]. It is the very local interstellar medium and magnetic field that combined with the solar wind determine the size and shape of the heliosphere [Baranov et al., 1981; Zank & Müller, 2003; Izmodenov et al., 2003]. Therefore observations of the boundaries of the heliosphere also constitute an indirect measurement of the local interstellar medium. More recent models of the heliosphere have been made which include IBEX measurements (such as interstellar magnetic field direction) as constraints [Pogorelov et al., 2011; Opher et al., 2011; Rathkiewicz et al., 2012]. In the present work we offer further IBEX measurements that can constrain models of the heliospheric interaction with the local interstellar medium.

Ionization processes erode the neutral interstellar wind on its passage through the heliosphere, and only a small fraction survives into the inner heliosphere. As a result, the composition of the wind at 1AU (where IBEX resides) is mostly helium, and measurements of the helium wind by IBEX give us our best determination of the local interstellar parameters [Möbius et al., 2012; Lee et al., 2012; Bzowski et al., 2012; McComas et al., 2012]. The hydrogen component was first observed by IBEX in 2009 as reported by [Moebius et al., 2009]. Later, a more detailed investigation of the Hydrogen component showed some variation over the first two years of operation [Saul et al., 2012]. The variation we report here over the last two years is much larger, and forms a strong test for global models of the heliosphere, in particular our models of ionization rates and radiation pressure.

**3) Longitudinal Structure of LISM H wind at 1 AU**

As the Earth (and IBEX) moves around the Sun, the instrument aperture views successively higher ecliptic longitudes. This change of pointing direction is not uniform, but is carried out by discrete orbital adjustments, which maintain the spacecraft attitude to maintain a nearly Sun pointing spin axis. Between the orbital adjustments, the orientation of IBEX's viewing is inertially fixed while the position of the spacecraft continues to move about the Sun, creating a "sawtooth" shape of the resulting time series [Saul et al., 2012]. The vertical discontinuities correspond to the orbital adjustments of the IBEX spacecraft attitude (Fig 2).

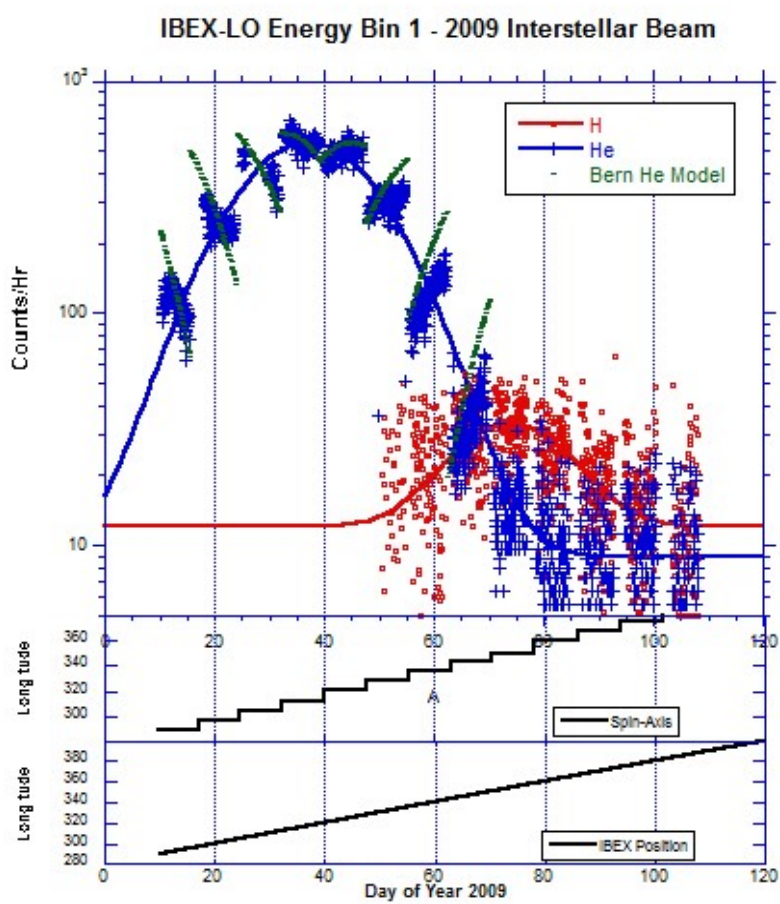

Figure 2. Overview plot for ~~qualitative~~ showing the hourly count rates in IBEX-Lo's first energy channel. Counts are separated into converted H atoms (red) and H atoms sputtered from the conversion surface by incident He (blue). A simulation (green) is based on the real pointing and position of the spacecraft (lower panels) Note the logarithmic scale. Gaussian fits are drawn to guide the eye.

For a colder distribution of neutral interstellar atoms, this sawtooth structure will be more pronounced, as there will be a larger difference between the flux from two nearby directions. For hydrogen, the sawtooth structure is barely visible (Figures 2 and 3). This is in part due to lower statistics but also suggests that the hydrogen distribution observed at 1AU appears hotter than the helium distribution, which serves as the reference for the physical conditions in the local interstellar medium [Möbius et al, 2012; Bzowski et al., 2012; McComas et al., 2012]. In Figure 3 we show the longitudinal profile of interstellar hydrogen observed by IBEX-Lo over the first four years of observation, where the spread in angular space is larger than for helium. To improve the statistics required to perform the separation of converted H from sputtered H, we use three-hour averages (Fig. 3) in contrast to the overview plot (Fig. 2).

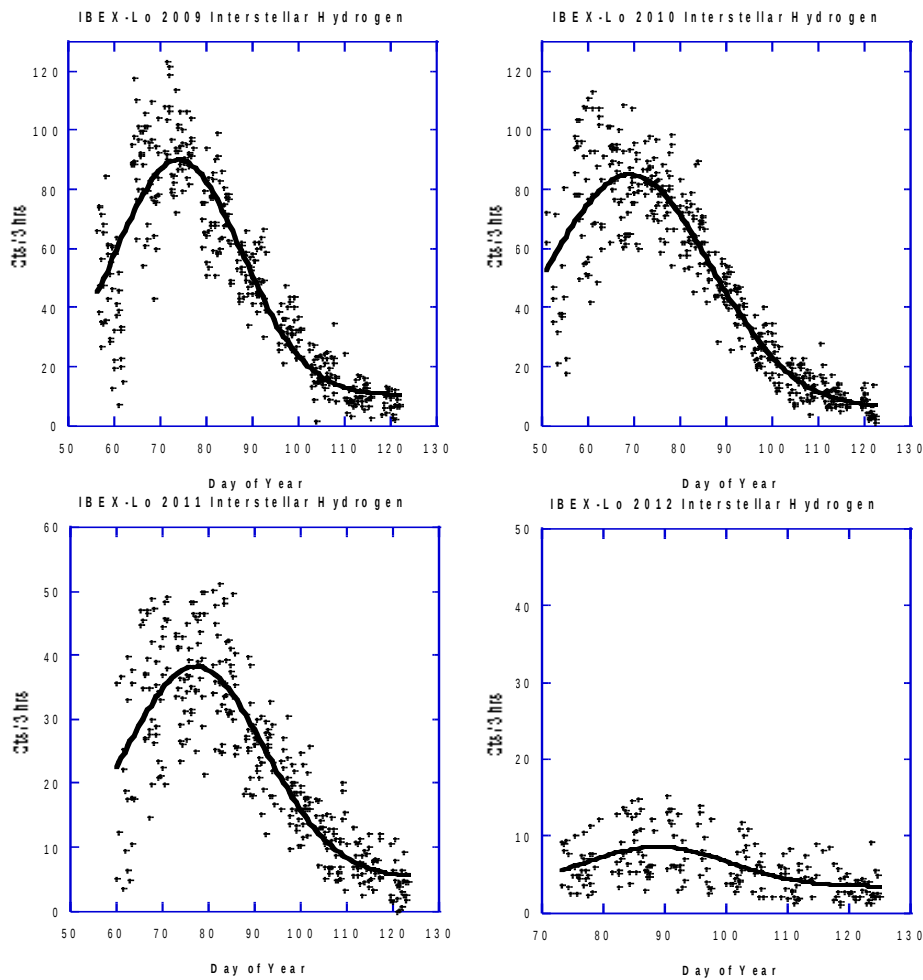

Figure 3: IBEX-Lo observations of interstellar hydrogen over the first four years of operation, shown as accumulated counts per 3 hours, as a function of Day of Year, for the lowest energy channel. Gaussian best fits are shown as solid lines; parameters with error bars are given in Table 1.

As IBEX observed the interstellar hydrogen in 2009 and 2010, the Sun was still in an extended solar minimum. The reduction in the LISM H flux observed over these years was quite small (see Fig. 3. and Table 1). However, as the Sun began to increase its activity and the extreme ultraviolet flux increased accordingly, leading directly to a dramatic decrease in the observed interstellar hydrogen. To derive the peak rate, the center position, and the width of the distribution we fit a Gaussian distribution and additive flat background. Table 1 provides the peak rate (in counts / 3 hrs) integrated over the entire range of heliospheric latitudes. The observed center of the H wind or the peak direction is given as the flow direction in heliospheric longitude. The background rate is included as a "best" parameter in the least squares fitting procedure, which is also listed in Table 1. The width is the standard deviation of the peak in degrees. It can be seen that the center of the peak moves to a later observation date with increasing solar activity. The position of the peak is in qualitative agreement with the result that the radiation pressure on hydrogen exceeds the gravitational force [Fahr et al., 1979; McComas et al., 2004]. The motion of this observed peak is in qualitative agreement with one driver: an increased radiation pressure as solar activity rises. We return to a more detailed discussion of comparing these observations to the modeled distribution of interstellar Hydrogen in the heliosphere in section 5.

|      | peak rate (cts) | background rate (cts) | Flow center (Ecl. Lat, Deg.) | width ($\sigma$) (Deg) |
|------|-----------------|-----------------------|------------------------------|------------------------|
| 2009 | 90.1 (1.8)      | 10.4 (1.6)            | 82.9 (0.3)                   | 19.4 (0.7)             |
| 2010 | 85.0 (1.7)      | 6.1 (1.7)             | 78.1 (0.4)                   | 25.0 (0.9)             |
| 2011 | 38.2 (1.1)      | 5.2 (1.1)             | 86.1 (0.5)                   | 21.4 (1.1)             |
| 2012 | 8.6 (0.6)       | 3.4 (0.5)             | 97.6 (1.2)                   | 16.5 (2.6)             |

Table 1. The best Gaussian fit to IBEX-Lo measurements of interstellar hydrogen flux. The flux is measured as counts per three hours in the first energy bin, while the center is given as the flow direction of the peak flux in heliospheric longitude. The width is the standard deviation of the distribution given in degrees. The background rate is calculated as a best fit from the data and so indicates not only real background but also deviation from the Gaussian fit. Statistical errors are given in parentheses.

**4) Latitudinal Profile of LISM H wind at Earth orbit**

IBEX is a Sun-pointed spinner and thus provides a latitude profile as a function of spin phase. The aperture of IBEX-Lo in the low angular resolution mode (used for these observations) is ~7 degrees full width half max [Fuselier et al., 2009]. We report here on the latitudinal profile of the interstellar H wind averaged over the peak orbit of H observation. The number of the IBEX orbit as well as the Gaussian fit characteristics are shown in Table 2.

By taking 2-degree bins in the spacecraft spin phase and summing over the good observations of the orbit at maximum LISM H viewing, we produce the latitude profiles shown in Figure 4. Again, these profiles show the reduction in H flux over the rising phase of the solar cycle. The differences in the center position (the flow direction in heliospheric latitude) are not as pronounced as those in heliospheric longitude. This is because the flow direction in latitude is close to the ecliptic plane and so radiation pressure does not have a chance to strongly deflect this component of the flow.

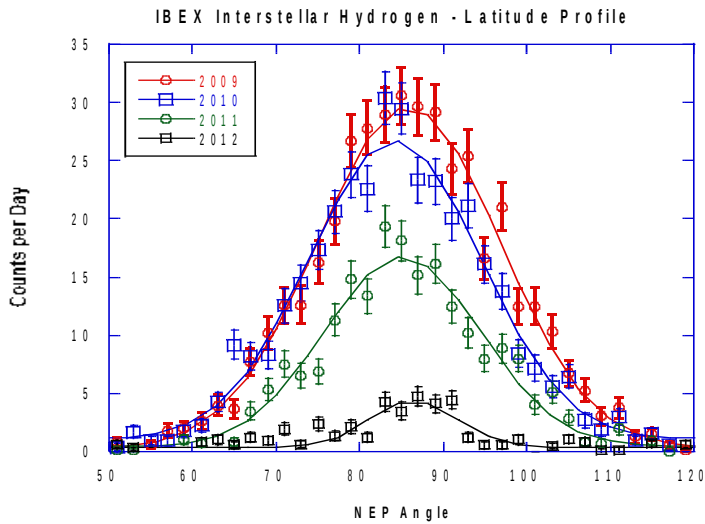

Figure 4: IBEX observations of interstellar hydrogen over the first four years of operation, shown as spin phase or latitudinal profiles as north ecliptic polar angle during the maximum LISM H flux orbits. Gaussian fits are shown as solid lines, parameters given in Table 2.

The width of the latitudinal profile also changes (Table 2), as the distribution appears to get cooler with increasing solar activity, a statistically significant variation. This may in part be due to filtering effects in the outer heliosphere as the higher energy and more direct trajectories have a higher survival probability in the ionizing field of the Sun. However the effects of filtering mechanisms on the distribution of H atoms at 1AU must be more fully modeled to quantitatively understand the implications.

|  | orbit | apogee | peak rate | center | width (σ) |
|---|---|---|---|---|---|
|  | (Num) | (Mon:D:H:Min) | (Cts) | (Ecl. Lat., Deg) | (Deg.) |
| **2009** | 23 | 03:30:14:33 | 29.43 (0.37) | -4.10 (0.15) | 15.47 (0.22) |
| **2010** | 71 | 03:31:12:55 | 25.76 (0.56) | -5.76 (0.25) | 14.55 (0.37) |
| **2011** | 119 | 03:31:16:28 | 16.32 (0.57) | -4.73 (0.57) | 13.27 (0.82) |
| **2012** | 161 | 04:02:16:12 | 4.04 (1.02) | -3.55 (1.52) | 7.36 (2.18) |

Table 2: IBEX-Lo observations of interstellar hydrogen over the first four years of operation, shown as spin phase or latitudinal profiles during the maximum LISM H flux orbits. The orbit number and date of apogee are given in the table as well as the peak rate in the angle bin (in counts per day) the flow direction center (in heliospheric latitude) and the standard deviation of the best fit Gaussian. Statistical errors are given in parentheses.

## 5) Discussion and Future Observations

Our observations show clearly that the interstellar hydrogen wind in the vicinity of the Earth has drastically reduced in the rising phase of the solar cycle. The precise amounts of this reduction, and the associated angular shifts in position (Tables 1 and 2), provide important information to test our knowledge of transport and loss of the interstellar atoms on their way in from the interstellar medium to Earth's orbit. While changes in the angular directions and flux are qualitatively consistent with those expected from the rising solar activity, the details lead to several questions: Once these observations are deconvolved with well-measured ionization rates and radiation pressure will they allow us to determine the density of protons in the heliosheath? Can we explain the observed "cooling" of the latitudinal angular distribution of the LISM H wind with velocity dependent filtration by ionizing alone? Do our ionization models account for these dramatic changes, and what do they predict for the coming solar maximum and the declining phase? Is the radiation pressure calculated based on direct observations of the Sun's Ly-$\alpha$ line profile and flux consistent with the radiation pressure needed to explain the interstellar hydrogen wind observations?

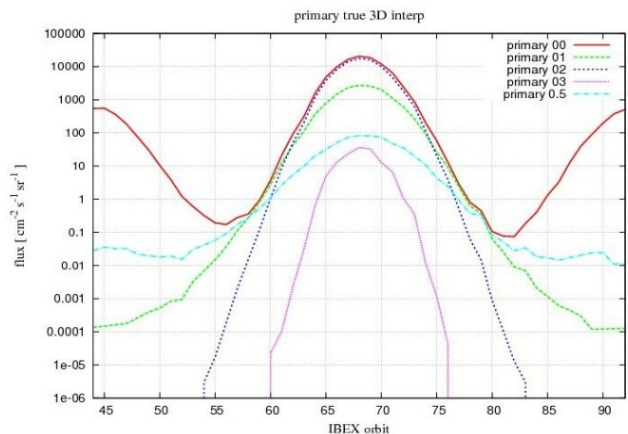

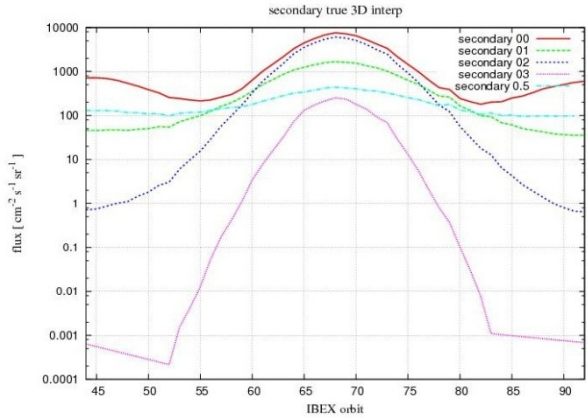

Figure 5. Model of expected H flux. Primary and secondary

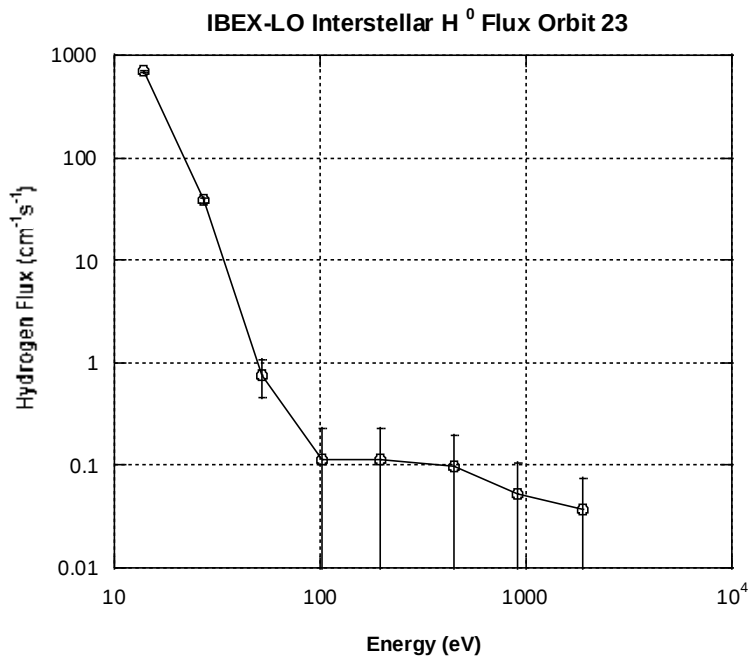

interstellar H populations based on simulations of the Warsaw Group [Bzowski et al. 2008; Bzowski et al. 2012]. The red line is the integral of all particles in the interstellar wind while the other lines show fluxes in the energy range of the first three IBEX-Lo energy bins

Figure 6. Flux of interstellar Hydrogen during orbit 23 is shown as measured at 1AU by IBEX-Lo (spacecraft reference frame). Omnidirectional background has been subtracted.

/ We have presented the interstellar hydrogen observed with IBEX-Lo in the lowest energy channel,

which is centered at 14 eV in the spacecraft frame of reference. This is near to the expected energy of unmodified interstellar hydrogen entering the heliosphere. While this is the energy channel where we observed the maximum H flux, there is also some hydrogen observed in energy channel 2, which is centered at 28 eV (Fig. 2). Understanding the small additional H flow at higher energies will need detailed modeling. We present here (Fig. 5) preliminary results of the Warsaw group to model these observations using the heliospheric parameters and ionization models which include heliospheric filtration presented in [Bzowski et al. 2012, Bzowski et al. 2008].

The most striking difference between these model results (Figure 5) and our observations is that they predict a peak signal in IBEX-Lo energy bin 2, whereas in our observations the peak is located in energy bin 1 (Figure 1 and Figure 6). The reason the models show the peak in bin 2 is in part due to heavy ionization loses experienced by lower energy Hydrogen entering the heliosphere. To emphasize this energy distribution we show in Figure 6 the same data from orbit 23 that was used to produce Figure 1. In this case we plot the flux of interstellar Hydrogen observed at the spacecraft rather than the count rate. We also do a background subtraction in which the average signal from the area of the sky not containing the LISM wind is subtracted from our signal (as carried out in Rodriguez et al. 2012). It can be seen that the higher energy background is removed and the enhancement in the low energy is accentuated after the instrument efficiencies are taken into account.

This energy discrepency could mean that some assumptions in the ionization model need new attention, including our estimates of Hydrogen densities in the outer heliosphere. Further modeling efforts underway are likely to test these assumptions against the time variability and the latitude distribution as well as the Ecliptic Longitude dependence of the interstellar H signal shown here.

   Finally, the nature of these observations in the rising solar cycle underline the importance of continued observations to separate the effects on the observed flow by the Sun from actual variations at the heliospheric interface. From our observations we can extrapolate over the rising solar activity and say that interstellar hydrogen will almost certainly not be visible during the solar maximum. Therefore we suggest an intense viewing strategy centered on the observing times available to IBEX during the falling solar cycle. Some improvements can be made to the observing strategy to increase our knowledge of interstellar hydrogen during that time. The simplest way to increase observing statistics will be to spend more time in the first energy bin of IBEX-Lo, rather than performing the nominal eight energy steps.

   In combination with direct ENA measurements from IBEX and in-situ measurements from Voyager, the neutral H wind distribution variations reported here add another way to test our knowledge of the outer heliosphere.

### 7. Acknowledgments

We acknowledge the entire IBEX team for their great work and dedication to this successful mission. Work carried out in the United States was supported by the IBEX mission, which is a part of NASA's Explorer program. The financial support of the Swiss National Foundation and hospitality and work from the University of Bern team is also acknowledged. Participants of a scientific meeting on the very local interstellar medium led by Dimitra Koutroumpa at the International Space Science Institute in Bern are acknowledged for useful discussion. Finally our reviewer is acknowledged for useful discussion and advice to improve this manuscript.

## References

Baranov, V. B., Ermakov, M. K., & Lebedev, M. G. (1981). A Three-Component Model of the Solar Wind - Interstellar Ineraction - Some Numerical Results. *Soviet Astron. Lett.*, *7*(3), 206.

Bzowsi, M., Kubiak, M. A., Moebius, E., Bochsler, P., Leonard, T., Heirtzler, D., & Kucharek, H. (2012). Neutral interstellar helium parameters based on IBEX-Lo observations and test particle calculations. *Astrophysical Journal Supplement Series*, *198*(12), 40.

Bzowski, M. (2008). Survival probability and energy modification of hydrogen Energetic Neutral Atoms on their way from the etermination shock to Earth orbit. *Astron. & Astrophys.*, *488*, 1057-1068.

Chassefière, E., Bertaux, J. L., & Sidis, V. (1986). Elastic collisions of solar wind protons with interstellar neutrals (H and He) inside the heliosphere: a new approach. *Astronomy and Astrophysics.*, *169*, 298-304.

Fahr. (1979). Interstellar Hydrogen Subject to a Net Repulsive Solar Force Field. *Astron. & Astrophys.*, *77*, 101-109.

Frisch, P. C., Grodnicki, L., & Welty, D. E. (2002). The velocity distribution of the nearest interstellar gas. *Astrophysical Journal*, *574*, 834-836.

Frisch, P. C., Redfield, S. & Slavin, J. D. The Interstellar Medium Surrounding the Sun. Annual Review of Astronomy and Astrophysics 49, 237–279 (2011).

Funsten, H. O., Allegrini, F., Bochsler, P., Dunn, G., Ellis, S., Everett, D., Fagan, M. J., et al. (2009). The Interstellar Boundary Explorer High Energy (IBEX-Hi) Neutral Atom Imager. *Space Science Reviews*, *146*, 75-103. Retrieved from 10.1007/s11214-009-9504-y

Fuselier, S. A., Bochsler, P., Chornay, D., Clark, G., Crew, G. B., Dunn, G., Ellis, S., et al. (2009). The IBEX-LO Sensor. *Space Sci Rev*, *146*(1-4), 117-147.

Izmodenov, V., Malama, Y. G., Gloeckler, G., & Geiss, J. (2003). Effects of Interstellar and Solar Wind Ionized Helium on the Interaction of the Solar Wind with the Local Interstellar Medium. *Astrophysical Journal*, *594*, L59-L62.

Katushkina, O. A., & Izmodenov, V. V. (2010). Effect of the heliospheric interface on the distribution of interstellar hydrogen atom inside the heliosphere. *Astronomy Letters*, *36*(4), 297-306.

Lallement, R., Raymond, J. R., & Vallerga, J. (2004). Diagnostics of the Local Interstellar Medium using particles and UV radiation. *Adv. Space Res.*, *34*(1), 46-52.

Lee, Martin A, Kucharek, H., Eberhard, M., Wu, X., Bzowski, M., & Mccomas, D. (2012). AN ANALYTICAL MODEL OF INTERSTELLAR GAS IN THE HELIOSPHERE TAILORED TO. *Astrophysical Journal Supplement Series*, *10*. doi:10.1088/0067-0049/198/2/10

McComas, D. J., Allegrini, F., Bochsler, P., & al., et. (2009). IBEX - Interstellar Boundary Explorer. *Space Sci Rev*, *146*(1-4), 11-33.

McComas, D. J., Schwadron, N., Crary, F. J., Elliot, H. A., Young, D. T., Gosling, J. T., Thomsen, M. F., et al. (2004). The interstellar hydrogen shadow: Observations of interstellar pickup ions beyond Jupiter. *Journal of Geophysical Research*, *109*, A02104.

McComas, D.J., D. Alexashov, M. Bzowski, H. Fahr, J. Heerikhuisen, V. Izmodenov, M.A. Lee, E. Moebius, N. Pogorelov, N.A. Schwadron, and G.P. Zank, The heliosphere's interstellar interaction: No bow shock, *Science*, 336, 1291, doi: 10.1126/science.1221054, May 2012.

Möbius, E., Bochsler, P., Bzowski, M., Heirtzler, D., Kubiak, M. A., Kucharek, H., Lee, M. A., et al. (2012). INTERSTELLAR GAS FLOW PARAMETERS DERIVED FROM INTERSTELLAR BOUNDARY EXPLORER-Lo OBSERVATIONS IN 2009 AND 2010: ANALYTICAL ANALYSIS. *Astrophysical Journal Supplement Series*, *11*. doi:10.1088/0067-0049/198/2/11

Möbius, E., Fuselier, S. A., Granoff, M. S., & al., et. (2008). Time-of-Flight Detector System of the IBEX-Lo Sensor with Low Background Performance for Heliospheric ENA Detection. (R. Caballero, Ed.)*30th International Cosmic Ray Conference*. Mérida.

Müller, Hans-Reinhard, Frisch, P. C., Florinski, V., & Zank, G. P. (2006). Heliospheric response to different possible interstellar environments. *Astrophysical Journal*, *647*, 1491-1505.

Opher, M. et al. Is the Magnetic Field in the Heliosheath Laminar or a Turbulent Sea of Bubbles? The Astrophysical Journal 734, 71 (2011).

Pogorelov, N. V. et al. Interstellar Boundary Explorer Measurements and Magnetic Field in the Vicinity of the Heliopause. The Astrophysical Journal 742, 104 (2011).

Ratkiewicz, R., Strumik, M. & Grygorczuk, J. the Effects of Local Interstellar Magnetic Field on Energetic Neutral Atom Sky Maps. The Astrophysical Journal 756, 3 (2012).

Redfield, S., & Linsky, J. L. (2008). The structure of the local interstellar medium. IV. Dynamics, morphology, physical properties, and implications of cloud-cloud interactions. *Astrophysical Journal*, *673*, 283-314.

Rodríguez M., D. F. et al. IBEX-Lo observations of energetic neutral hydrogen atoms originating from the lunar surface. Planetary and Space Science 60, 297–303 (2012).Saul, Lukas, Wurz, P., Rodriguez, D., Eberhard, M., Schwadron, N., Kucharek, H., Leonard, T., et al. (2012). Local interstellar neutral hydrogen sampled in situ by IBEX. *Astrophysical Journal Supplement Series*, *14*, 2-9. doi:10.1088/0067-0049/198/2/14

Scheer, J.A., M. Wieser, P. Wurz, P. Bochsler, E. Hertzberg, S.A. Fuselier, F.A. Koeck, R.J. Nemanich, and M. Schleberger, **"Conversion Surfaces for Neutral Particle Imaging Detectors,"** Adv. Space Res. 38 (2006), 664-671.

Vallerga, J., Lallement, R., Lemoine, M., Dalaudier, F., & McMullin, D. (2004). EUVE observations of the helium glow: Interstellar and solar parameters. *Astron. & Astrophys.*, *426*, 855-865.

Wurz, P, Fuselier, S. A., Möbius, E., Funsten, H. O., Brandt, P. C., Allegrini, F., Ghielmetti, A., et al. (2009). IBEX Backgrounds and Signal-to-Noise Ratio. *Space Sci Rev*, *146*(1-4), 173-206.

Wurz, P, Saul, L., Scheer, J. A., Möbius, E., Kucharek, H., & Fuselier, S. A. (2008). Negative Helium Generation upon Surface Scattering - Application to Space Science. *Jou. Appl. Phys.*, *103*.

Wurz, P., **Detection of Energetic Neutral Particles,** in *The Outer Heliosphere: Beyond the Planets*, (eds. K. Scherer, H. Fichtner, and E. Marsch), Copernicus Gesellschaft e.V., Katlenburg-Lindau, Germany, (2000), 251–288.

Wurz, P., J. Scheer, and M. Wieser, **"Particle scattering off surfaces: application in space science,"** e-Jou. Surf. Science Nanotechn. 4 (2006) 394-400.

Zank, G P, & Müller, H.-R. (2003). The dynamical heliosphere. *Journal of Geophysical Research*, *108*(A6), 1240.